\documentclass[notitlepage,article,showpacs,preprintnumbers,amsmath,amssymb,prb,superscriptaddress,footinbib]{revtex4-1}

\usepackage{graphicx}% Include figure files
\usepackage{dcolumn}% Align table columns on decimal point
\usepackage{bm}% bold math
\usepackage{epstopdf}
\usepackage{float}
\usepackage[normalem]{ulem}
\DeclareGraphicsExtensions{.pdf,.eps,.png,.jpg,.mps} 

\usepackage{color, soul}

\usepackage{xcolor}

\definecolor{lightcyan}{rgb}{.6,.8,1}

\begin{document}

\title{Dynamical decoupling of interacting dipolar spin ensembles}
\author{Evan S. Petersen}
\affiliation{Department of Electrical Engineering, Princeton University}
\author{A. M. Tyryshkin}
\affiliation{Department of Electrical Engineering, Princeton University}

\author{K. M. Itoh}
\affiliation{School of Fundamental Science and Technology, Keio University}

\author{Joel W. Ager}
\affiliation{Lawrence Berkeley National Laboratory}

\author{H. Riemann}
\affiliation{Leibniz-Institut für Kristallzüchtung}
\author{N. V. Abrosimov}
\affiliation{Leibniz-Institut für Kristallzüchtung}
\author{P. Becker}
\affiliation{PTB Braunschweig}
\author{H.-J. Pohl}
\affiliation{VITCON Projectconsult GmbH}

\author{M. L. W. Thewalt}
\affiliation{Department of Physics, Simon Fraser University}

\author{S. A. Lyon}
 \affiliation{Department of Electrical Engineering, Princeton University}

\begin{abstract}
    We demonstrate that CPMG and XYXY decoupling sequences with non-ideal $\pi$ pulses can reduce dipolar interactions between spins of the same species in solids. 
    Our simulations of pulsed electron spin resonance (ESR) experiments show that $\pi$ rotations with small ($<$~10\%) imperfections refocus instantaneous diffusion. 
    Here, the intractable N-body problem of interacting dipoles is approximated by the average evolution of a single spin in a changing mean field. 
    These calculations agree well with experiments and do not require powerful hardware. 
    Our results add to past attempts to explain similar phenomena in solid state nuclear magnetic resonance (NMR).
    Although the fundamental physics of NMR are similar to ESR, the larger linewidths in ESR and stronger dipolar interactions between electron spins compared to nuclear spins preclude drawing conclusions from NMR studies alone.
    For bulk spins, we also find that using XYXY results in less inflation of the deduced echo decay times as compared to decays obtained with CPMG.
\end{abstract}

\maketitle

Dynamical decoupling has been used for canceling interactions and extending coherence in nuclear magnetic resonance (NMR)\cite{cpmg1958, waugh1968_2, shaka1987} and more recently for counteracting noise on qubits.\cite{Ryan2010, Green2013, Suter2016} 
Decoupling sequences comprised of repeated $\pi$ pulses, such as CPMG\cite{Carr1954, cpmg1958} and XYXY,\cite{Gullion1990, ZhiHui2012} are used to obtain long spin qubit coherence times.\cite{Steger2012,Saeedi2013,Muhonen2014} 
However, the performance of these sequences on ensembles of interacting spins is not well understood.\cite{Barrett2007, Barrett2008, Tyryshkin2010, ZhiHui2012, Levstein2012, Walls2014, Leskes2015}
Furthermore, the existing analysis is predominantly for NMR,\cite{Barrett2007, Barrett2008, Levstein2012, Walls2014, Leskes2015} while related effects for electron spin resonance (ESR) are largely unexplored.
Using simulations of pulsed ESR, we find that sequences with slightly ($<$~10\%) imperfect $\pi$ pulses suppress decoherence from dipole-dipole interactions (specifically instantaneous diffusion), a mechanism absent from single spin measurements but common in bulk crystals. 
Past attempts to model these interactions have been limited due to the problem's N-body nature.\cite{Barrett2007, Barrett2008, Walls2014, Leskes2015} 
We partially circumvent that limitation by approximating instantaneous diffusion as a mean field, and also model imperfect $\pi$ pulses and global magnetic field noise. 
These simulations can be run quickly and agree well with experiments.

The simplest form of decoupling is the Hahn echo,\cite{Hahn1950} which refocuses spin evolution from static magnetic fields. By repeating this $\tau$-$\pi$-$\tau$ sequence many times with progressively shorter delays, $\tau$, spins are decoupled from dynamic fields changing slowly compared to $\tau$.\cite{Viola1999_2}
One popular implementation of this repetition is CPMG,\cite{cpmg1958} which always rotates spins about the same axis. In exchange for its simplicity, CPMG is only capable of preserving one specific spin state.
Sequences like XYXY\cite{Gullion1990} trade off simplicity to preserve all spin states, alternating $\pi$ rotations about different axes in varying permutations.\cite{Gullion1990, Viola1999_2,Tyryshkin2010,ZhiHui2012}

Non-ideal $\pi$ rotations cause unintended consequences when using decoupling sequences.\cite{Barrett2007, Barrett2008, Tyryshkin2010, ZhiHui2012, Levstein2012, Walls2014, Leskes2015}
Multiple NMR studies suggest non-ideal pulses in CPMG artificially elongate ensemble spin echo decays, with potential causes including spin-spin interactions during pulses,\cite{Barrett2007, Barrett2008} stimulated echoes,\cite{Levstein2012} spin-locking,\cite{Walls2014} and the reduction of effective dipolar interactions.\cite{Leskes2015}
Some of these possibilities imply that ensemble measurements using CPMG can misrepresent the average coherence time of individual spins, and suggest that other sequences be investigated.
\begin{figure}
\centering
\includegraphics[scale=.55,keepaspectratio]{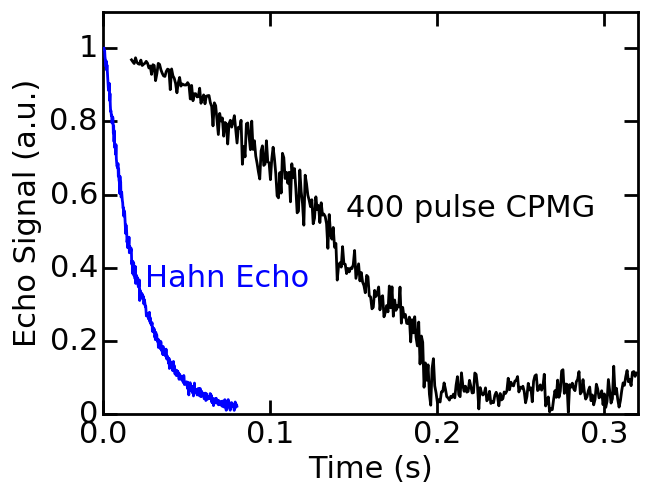}
\caption{\label{fig:cpmgVsHahn} (color online) Magnitude detected 400 pulse CPMG (black) and Hahn echo (blue) decays of donor electron spins from a $1.2 \times 10^{14}~\text{P/cm}^3$ $^{28}$Si crystal at 4.8~K. }
\end{figure}

To evaluate how dynamical decoupling affects coherence experiments, we study spins in a crystal where the cause of decoherence is already known.
We compare measurements and simulations of echo decays with and without dynamical decoupling.
$^{31}$P donor electron spin coherence in a $^{28}$Si crystal with 50ppm $^{29}$Si and $1.2~\times10^{14} \text{P/cm}^3$ was previously found to be limited by instantaneous diffusion.\cite{Tyryshkin2011}
Instantaneous diffusion occurs when a spin's resonant frequency changes following a $\pi$ pulse due to the reversed dipolar fields from surrounding spins rotated by the same pulse.\cite{Klauder1962, Salikhov1981}
From an operator perspective, the dipolar interaction commutes with $\pi$ rotations on ensemble spins and is not refocused.
Therefore, in the absence of pulse errors or other sources of decoherence, Hahn echo and dynamical decoupling experiments should produce identical echo decays in our crystal. 

Pulsed ESR experiments are performed using a Bruker Elexsys E580 spectrometer.
Measurements were made at temperatures between 1.7 and 4.8~K with an applied magnetic field oriented along the [100] crystal axis.
To counteract magnetic field noise (which dephases all spins equally in our system), we measure echoes with magnitude detection.\cite{Tyryshkin2003}
In Figure 1, we plot magnitude detected measurements of both a Hahn echo decay and an echo decay obtained with a 400 pulse CPMG sequence.
The large increase in decay time when using CPMG suggests that the two methods affect instantaneous diffusion differently.
Our simulations below show that errors in the axis and angle of rotations, which break the commutation between pulses and dipole-dipole coupling, are responsible for the elongated decay.

\begin{figure}
\centering
\includegraphics[scale=.55,keepaspectratio]{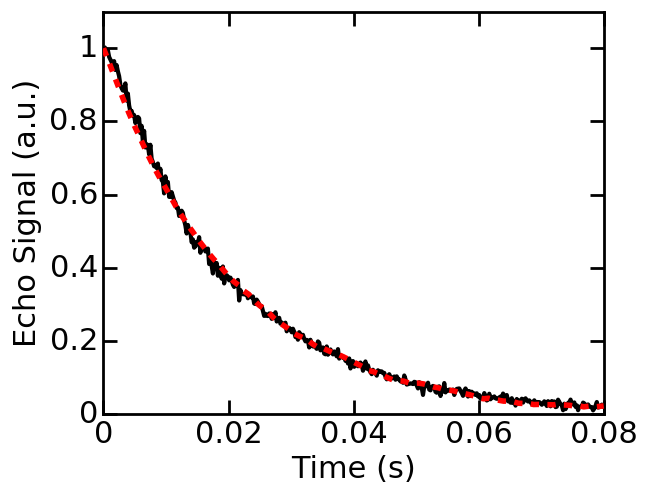}
\caption{\label{fig:hahnDecay} (color online) Measured (black) and simulated (red dashes) Hahn echo decays. Broadening\cite{Kittel1953} from instantaneous diffusion is calculated for a donor density of $1.2 \times 10^{14} \text{P/cm}^3$.}
\end{figure}

The evolution of spins due to instantaneous diffusion during a dynamical decoupling sequence is an N-body problem, restricting exact numerical methods to a handful of spins.\cite{Leskes2015, Walls2014} 
Even neglecting pulses and considering only spin-spin interactions, the exact diagonalization of a Hamiltonian has been limited to $\sim$~40 spins and only possible for specific lattice configurations.\cite{Richter2010}
However, our particular problem can be approximated as a calculation of a single \textit{central} spin's evolution while treating the dipolar interaction to other spins as a mean magnetic field. The magnitude of the magnetic field is described as inhomogeneous dipolar broadening, previously shown to be Lorentzian with half width at half maximum (HWHM):\cite{Kittel1953}
\begin{eqnarray}
\text{HWHM} = 5.3 \times \frac{\gamma^2 \mu_0 \hbar n}{4 \pi}
\end{eqnarray}
\noindent for spin gyromagnetic ratio $\gamma$, vacuum permeability $\mu_0$, reduced Planck constant $\hbar$, and spin density $n$. The field's contribution to a spin's resonance frequency switches sign following a $\pi$ rotation. Our $\pi$ pulses are tuned to one branch of the hyperfine splitting of the spins, so $n$ is taken to be half of the donor density (evolution from other spins' fields is refocused by the pulse). The central spin evolution operator describing a Hahn echo experiment with dipole-dipole interactions is written:
\begin{eqnarray}
&U = U_{\mathcal{H}-} R_{\pi,X} U_{\mathcal{H}+}\\
&U_{\mathcal{H}\pm} = e^{-i \tau S_Z (\omega_Z \pm \omega_{bath})}
\label{eq:upm}
\end{eqnarray}
\noindent where $R_{\pi,X}$ is the operator for a $\pi$ rotation about the x axis, $\tau$ is the delay time before and after the $\pi$ pulse, $S_Z$ is the spin-1/2 Z operator, $\omega_z$ is the spin's Zeeman frequency, and $\omega_{bath}$ is the shift in resonance frequency due to a bath of dipoles. The ensemble echo signal is found by averaging single spin evolutions under $U_\mathcal{H}$ for different values of $\omega_Z$ and $\omega_{bath}$. Each spin starts in a superposition state represented by density matrix $\rho_{init}$ (e.g. $\rho_{init}=S_X$). After applying $U$ to $\rho_{init}$, we calculate the projection of the resulting matrix onto $\rho_{init}$, corresponding to the echo strength. The projection onto the perpendicular state in the x-y plane (e.g. $S_Y$ for $\rho_{init}=S_X$) is also calculated for magnitude detection. In each evolution, $\omega_Z$ is taken from a distribution previously determined for our experimental setup by Wang et al.\cite{ZhiHui2012} and $\omega_{bath}$ is picked from the distribution derived by Kittel and Abrahams.\cite{Kittel1953} Numerical calculations of the echo decay are shown as red dashes in Fig.~\ref{fig:hahnDecay}. This model accurately reproduces our measured echo decay, shown in black.  

\begin{figure}
\centering
\includegraphics[scale=.55, keepaspectratio]{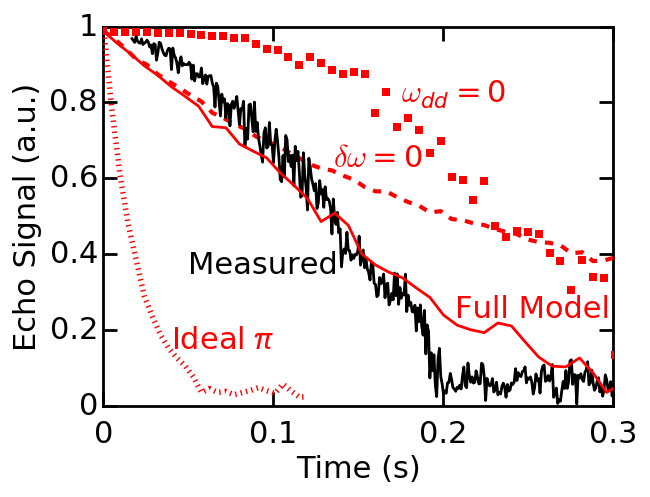}
\caption{\label{fig:cpmgModel} (color online) Measured (black) and simulated (red) echo decays from a 400 pulse CPMG sequence at 4.8~K. 
    With ideal $\pi$ rotations (dotted line), our model decays at the same rate as a Hahn echo. 
    When pulse errors are introduced without field noise noise but including instantaneous diffusion ($\delta\omega=0$, red dashes), the echo decay time is extended at long times. 
    Alternitavely, the echo decay from global field noise and no dipole-dipole interactions ($\omega_{dd}=0$, red squares) is slower. 
    Including pulse errors, instantaneous diffusion, and field noise (Full Model) produces the solid red line.}
\end{figure}

Pulse-to-pulse variations (errors) do not allow one to reproduce a dynamical decoupling sequence by simply repeating $U$. The effect of ignoring errors is shown in Fig.~\ref{fig:cpmgModel} with a simulated echo decay for a 400 pulse CPMG sequence, obtained by applying $U$ 400 times (red dotted line labeled ``Ideal $\pi$''). This simulated decay matches the Hahn echo decay in Fig.~\ref{fig:hahnDecay}, while the measured decay from a 400 pulse sequence, shown in black in Fig.~\ref{fig:cpmgModel}, is clearly longer. Our simulations show that the pulse errors modify the dipole-dipole coupling between spins. The Hamiltonian, $\mathcal{H}_{dd}$, for two dipole-dipole~coupled~spins~is:
\begin{subequations}
\begin{align}
&\mathcal{H}_{dd} = \hbar\omega_{dd_{i,j}}(S_{Z_i}S_{Z_j} - \frac{1}{4}\left(S_i^+S_j^- + S_i^-S_j^+\right))\\
&\omega_{dd_{i,j}}=\frac{\zeta_{i,j}\left(1-3cos^2\left(\theta\right)\right)}{r^3}\\
&\zeta_{i,j} =\gamma_i\gamma_j\mu_0\hbar/4\pi
\end{align}
\end{subequations}
\noindent where $\gamma_i$ is the gyromagnetic ratio of spin $i$, $\theta$ is the angle between spins relative to the applied magnetic field, and $S_i^+$ and $S_i^-$ are the raising and lowering operators for spin $i$. The $S_{Z_i}S_{Z_j}$ term causes instantaneous diffusion and the $S_i^+S_j^- + S_i^-S_j^+$ term is responsible for spin flip-flops. Spin flip-flops are slow compared to the timescales of all but the longest decoupling sequences in this work\cite{Tyryshkin2011, Petersen2017} and are therefore ignored in our calculations. As mentioned above, the $S_Z S_Z$ term commutes with a $\pi$ rotation of both spins ($R_{\pi,X,i}R_{\pi,Y,j}$), so it is not surprising that our first simulation of a 400 pulse CPMG sequence simply recreated the Hahn echo measurement.

Our simulations must include effects from pulse errors on both the central and bath spins, resulting in elongated decays (as shown in red dashes in Fig. \ref{fig:cpmgModel} for 400 pulse CPMG). Wang \textit{et al.}\cite{ZhiHui2012} previously describe an effective means of modeling pulse errors on a central spin by replacing the $R_{\pi,X}$ rotation operators with ones including angle and axis of rotation errors:
\begin{eqnarray}
\begin{split}
&R_{\pi,X} = \exp{\left(\left(\pi + \alpha\right) \left[\sqrt{1-n_z}S_X + n_z S_Z\right]\right)}
\end{split}
\end{eqnarray}
\noindent where $\alpha$ is the angle of rotation error and $n_z$ is the axis of rotation error. These errors are taken from distributions previously measured for our system.\cite{ZhiHui2012,Tyryshkin2010} 

Importing pulse errors into the spin bath is accomplished by modifying the mean field after each pulse. The free evolution of a central spin surrounded by bath spins is a superposition of evolutions under fields from all bath polarizations. Pulse errors introduce an additional layer of superposition - between evolutions under fields from bath spins that have and have not flipped following each pulse. In principle one could calculate the sequences of central spin evolution under each bath configuration throughout an experiment and determine a final state. That calculation is intractable, but for our small ($<$~10\%) pulse erors an average final state projection converges within 10,000 configurations. We obtain a mean field from the bath spins and recalculate that field after each pulse probabilistically flips each spin. Here we use the probability $P_{\text{flip}}$ of measuring a spin ($\uparrow$ or $\downarrow$) as flipped by a non-ideal $\pi$ pulse:
\begin{eqnarray}
P_{\text{flip}} = (1 - n_z^2) \cos^2{\left(\frac{\alpha}{2}\right)}.
\label{eq:pFlipModel}
\end{eqnarray} 
In our simulations the positions of bath spins are also averaged over to account for local variations in a dilute crystal.

With pulse errors included, simulated echo decays from instantaneous diffusion alone were longer than in experiments. This difference can be understood by considering decoherence from global field fluctuations (shown as red squares in Fig. \ref{fig:cpmgModel} for 400 pulse CPMG), determined by noise measurements in our system.\cite{Asfaw2015} Here, the $U_\mathcal{H}$ operator is replaced with a Trotter expansion containing $N$ operators, each including a frequency shift from noise integrated over time step $\tau/N$. For long CPMG sequences, magnitude detection can no longer counteract global field noise since CPMG only preserves a single spin axis, making echo decay from the field noise alone significant.

The final form of our model's free evolution operator, including instantaneous diffusion and field noise, is:
\begin{eqnarray}
U_\mathcal{H}(\tau) = \prod_n^{N}{U_{\mathcal{H},n}}(\tau) = \prod_n^{N}{\exp{(i \frac{\tau}{N} (\omega_Z + \omega_{bath} + \delta\omega(n,\tau/N)) S_Z)}}
\label{eq:finalU}
\end{eqnarray}
\noindent where $\delta\omega(n,\tau/N)$ is the frequency shift from noise in the $n$'th operator for evolution over time step $\tau/N$ and $\omega_{bath}$ is recalculated whenever a $\pi$ pulse is applied using Eq. \ref{eq:pFlipModel}. Simulated CPMG echo decays are significantly improved in this model, as shown by the solid red line (Full Model) in Fig.~\ref{fig:cpmgModel} for a 400 pulse CPMG sequence. 

\begin{figure}
\centering
\includegraphics[scale=.52,keepaspectratio]{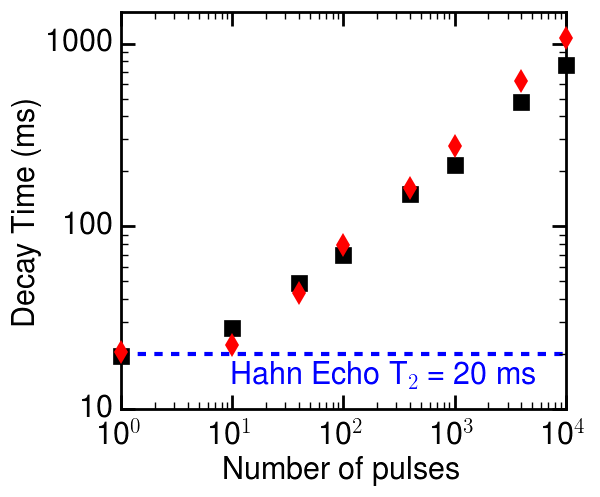}
\caption{\label{fig:cpmgNumber} (color online)  Measured (black squares) and simulated (red diamonds) CPMG echo decay times for a $^{28}$Si crystal doped with $1.2 \times 10^{14} \text{P/cm}^3$ at 4.8~K.}
\end{figure}

The decay times from CPMG sequences up to ten thousand pulses long, both measured and simulated, are shown in Fig.~\ref{fig:cpmgNumber}. For long CPMG sequences ($>1000$~pulses), simulated echo decays are slower than seen in experiments. The differences in decay times likely arise from flip-flops, absent in our model, occurring over time-spans comparable to these longer experiments.\cite{Tyryshkin2011} With the appropriate noise spectrum, these flip-flops could be included as an additional local field noise in Eq.~\ref{eq:finalU}.

There has been debate over what is measured by echo decays obtained from decoupling sequences composed of periodic $\pi$ pulses.
This work has largely been associated with NMR.\cite{Barrett2007, Barrett2008, Levstein2012, Walls2014, Leskes2015} 
Analytical treatments either consider only small numbers of spins,\cite{Barrett2007, Barrett2008, Leskes2015} neglect spin-spin interactions entirely,\cite{Levstein2012} or utilize a Magnus expansion to obtain an average Hamiltonian theory (AHT).\cite{Waugh1968, Walls2014}
The interpretation of ESR cannot in general use AHT because the product of linewidth and pulse delays is almost always $\gg1$. 
In this situation there is no basis to expect the Magnus expansion approach to be valid.\cite{Moan2008}

Earlier studies by Li \textit{et al.}\cite{Barrett2007, Barrett2008} conclude that dipole-dipole interactions between spins during the application of pulses result in extended $T_2$'s. 
However, Franzoni \textit{et al}\cite{Levstein2012} find these interactions have a negligible effect, and $T_2$'s are extended because stimulated echoes, whose lifetimes are elongated by $T_1$, are measured alongside Hahn echoes.
Ridge \textit{et al}\cite{Walls2014} also conclude that dipole-dipole interactions during pulses do not matter, and that stimulated echoes contribute to decay measurements.
However, they could not determine whether stimulated echoes are extended by $T_1$. 
Their N-body simulations, which depend on AHT, diverge from experiments with large inter-pulse spacings, suggesting a need for higher order corrections. 
Leskes and Grey\cite{Leskes2015} instead use Floquet methods for an N-body calculation, finding that dipole-dipole terms in the Floquet Hamiltonian are reduced by imperfections in $\pi$ pulses. 
They also find that contributions from stimulated echoes at the end of a CPMG sequence are negligible, and that there is no difference between their models that do or do not include them.

We treat N-body phenomena as a changing magnetic field and can have long times between pulses. 
Our results suggest that the $S_ZS_Z$ dipolar interaction is suppressed by non-ideal pulses, leading to longer echo decays. 
Separately, we investigated Franzoni \textit{et al.}'s suggestion \cite{Levstein2012} of mixing between $T_1$ and $T_2$ by including pre-determined exponential $T_1$ and $T_2$ decays in our models. 
Some large $\alpha$ and $n_z$ values can result in mixing, but $T_2$ is rarely inflated by more than 10\% for our pulse errors and becomes negligible after averaging. 
This holds true not only for ensembles but also for single spin experiments using CPMG.\cite{Muhonen2014, Laucht2016, Freer2017}
\begin{figure}
\centering
\includegraphics[scale=.52,keepaspectratio]{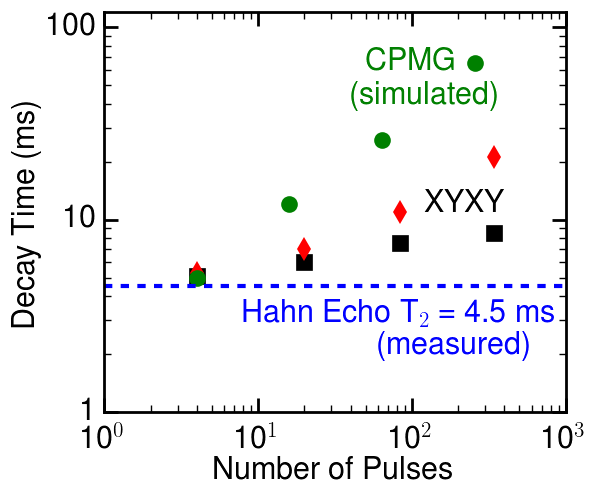}
    \caption{\label{fig:xyPlots} (color online) Experimental and simulated concatenated XYXY echo decay times as well as CPMG simulations for a $5 \times 10^{14}~\text{P/cm}^3$ doped $^{28}$Si crystal. Experimental data are shown in black squares (from [\onlinecite{Tyryshkin2010}]) and decay times from our model are shown in red diamonds (for XYXY) and green circles (for CPMG). The Hahn echo decay time, which was identical in both experiment and our model, is drawn as a reference with a dashed blue line.}
\end{figure}

XYXY\cite{Gullion1990, Viola1999_2,Tyryshkin2010,ZhiHui2012} can protect all components of a spin state,\cite{Tyryshkin2010,ZhiHui2012} while CPMG protects only one. Measurements with XYXY have substantially smaller increases in echo decay times. Concatenated asymmetric XYXY sequences were previously tested in our spectrometer on a $^{28}$Si crystal with 800ppm of $^{29}$Si and doped with $5 \times 10^{14}~\text{P/cm}^3$.\cite{Tyryshkin2010} The zeroth level concatenation of XYXY is a 4-pulse $\tau-\pi_x-\tau-\pi_y-\tau-\pi_x-\tau-\pi_y$ sequence, with higher concatenation levels nesting this sequence within its $\tau$ delays. We re-plot the data from ref.~\onlinecite{Tyryshkin2010} in black squares in Fig.~\ref{fig:xyPlots} alongside simulations using our model in red diamonds. Like the measured data, our model produces longer decay times for increasing levels of concatenation, but to a greater extent. Simulated decay times for CPMG sequences, shown in green circles in Fig.~\ref{fig:xyPlots}, are consistently longer than in comparable XYXY experiments. Even though XYXY at concatenation level 3 has 340 pulses, the increase in decay time (1.9$\times$ in experiment and 4.7$\times$ in our model) is much smaller than the 14$\times$ increase predicted by our model for a 256 pulse CPMG sequence. This result suggests that pulse errors in the XYXY approach do not suppress instantaneous diffusion as much as in CPMG measurements.

In conclusion, we can efficiently model dynamical decoupling sequences composed of $\pi$ pulses acting on ensembles of interacting spins. 
Our approach transforms the N-body problem of dipole-dipole coupled spins into a single spin calculation with a changing magnetic field. 
The resulting simulated decays are in agreement with experiments, and predict decay times within a factor of 3 of measured values. 
They also reproduce increasing T$_2$ times measured by CPMG sequences. 
These longer T$_2$ times are caused by pulse errors, which suppress dipole-dipole interactions in the form of instantaneous diffusion. 
This ensemble mechanism is not present in non-interacting spin experiments and we do not expect these extra complications to arise when using dynamical decoupling for very dilute or single spins assuming the pulse errors are sufficiently small. 
However, for measurements on bulk spins, applying XYXY instead of CPMG leads to less artificial inflation of the deduced echo decay times.
\bibliography{decouplingPapers}
\end{document}